\documentclass[10pt,conference]{IEEEtran}
\usepackage{amsmath,amsfonts}
\usepackage{algorithmic}
\usepackage{algorithm}
\usepackage{array}
\usepackage[caption=false,font=normalsize,labelfont=sf,textfont=sf]{subfig}
\usepackage{textcomp}
\usepackage{stfloats}
\usepackage{url}
\usepackage{verbatim}
\usepackage{graphicx}
\usepackage{cite}
\usepackage{gensymb}

\usepackage{multirow, booktabs, hhline}
\usepackage{color,soul}
\setulcolor{red}
\sethlcolor{yellow}
\usepackage{xcolor}

\renewcommand\hl[1]{#1} 
\usepackage{enumitem}
\setlist[itemize]{leftmargin=20mm}

\usepackage{tikz}
\usepackage{textcomp}
\usepackage{hyperref}
\usepackage{lipsum}

\newcommand\copyrighttext{%
  \footnotesize This paper has been accepted by the IEEE/IFIP Network Operations and Management Symposium (NOMS), 8–12 May 2023, Miami, FL, USA.\\ \copyright 2023 IEEE. Personal use of this material is permitted. Permission from IEEE must be obtained for all other uses, in any current or future media, including reprinting/republishing this material for advertising or promotional purposes, creating new collective works, for resale or redistribution to servers or lists, or reuse of any copyrighted component of this work in other works.}
\newcommand\copyrightnotice{%
\begin{tikzpicture}[remember picture,overlay]
\node[anchor=south,yshift=10pt] at (current page.south) {\fbox{\parbox{\dimexpr\textwidth-\fboxsep-\fboxrule\relax}{\copyrighttext}}};
\end{tikzpicture}%
}
\begin{document}
\bstctlcite{IEEEexample:BSTcontrol}

\title{SatAIOps: Revamping the Full Life-Cycle Satellite Network Operations}

\author{\IEEEauthorblockN{Peng Hu}
\IEEEauthorblockA{
\IEEEauthorblockA{\textit{Digital Technologies Research Center, National Research Council of Canada}
}
\IEEEauthorblockA{\textit{1200 Montreal Road, Ottawa, ON K1A 0R6. Canada}
}\\
}
\IEEEauthorblockA{ \vspace{-10pt}
\IEEEauthorblockA{\textit{The Cheriton School of Computer Science, University of Waterloo}
\IEEEauthorblockA{\textit{200 University Ave W., Waterloo, ON N2L 3G1. Canada}}}
peng.hu@nrc-cnrc.gc.ca
}\vspace{-20pt}
}

\markboth{Journal of \LaTeX\ Class Files,~Vol.~14, No.~8, August~2021}%
{Shell \MakeLowercase{\textit{et al.}}: A Sample Article Using IEEEtran.cls for IEEE Journals}


\maketitle
\copyrightnotice
\begin{abstract}
Recently advanced non-geostationary (NGSO) satellite networks represented by large constellations and advanced payloads provide great promises for enabling high-quality Internet connectivity to any place on Earth. However, the traditional approach to satellite operations cannot address the new challenges in the NGSO satellite networks imposed by the significant increase in complexity, security, resilience, and environmental concerns. Therefore, a reliable, sustainable, and efficient approach is required for the entire life-cycle of satellite network operations. This paper provides a timely response to the new challenges and proposes a novel approach called ``SatAIOps'' as an overall solution. Through our discussion on the current challenges of the advanced satellite networks, SatAIOps and its functional modules in the entire life-cycle of satellites are proposed, with some example technologies given. SatAIOps provides a new perspective for addressing operational challenges with trustworthy and responsible AI technologies. It enables a new framework for evolving and collaborative efforts from research and industry communities. 
\end{abstract}


\section{Introduction}
With the increasing launches of non-geostationary (NGSO) satellites, our future Internet infrastructure and critical systems such as telecommunications, transportation, financial services, Earth observation, and defence will rely on space-based systems. These systems include various entities, such as satellite networks, ground segments, and high-altitude platform stations (HAPS). Although the proliferated low-Earth orbit (LEO) satellites in large constellations have shed light on a high-quality Internet, the traditional satellite operational approach has not responded well to the challenges imposed by the fast-growing space assets. A new approach is much needed to provide \hl{reliable, efficient, and sustainable operations} for satellite networks. This work first discusses the challenges based on the current satellite operations in Section II. The proposed SatAIOps approach with its functional modules, fundamental services, and application domains are presented in Section III, followed by the conclusion made in Section IV. 

\section{Challenges in Current Satellite Operations}

The current satellite operations follow the traditional approach weighted on a manual process, centered on individual spacecraft, and focused on a few activities in a spacecraft's life cycle. For example, as shown in the highlighted blocks in Fig. \ref{Fig:SatAIOps_process}, the quality assurance \cite{Gerard2020} is handled in the \textit{build} phase when a satellite is manufactured with various hardware components. In the \textit{operate} phase, the management/maintenance through telemetry, tracking, and control (TT\&C), mission assurance, task management, and decommission sequences. These traditional functions need to be expanded and updated to respond to the new operational challenges.

\subsection{Efficiency Challenges in Satellite Operations}
\begin{figure}[ht]
    \vspace{-10pt}
    \centering
    \includegraphics[width=0.83\linewidth]{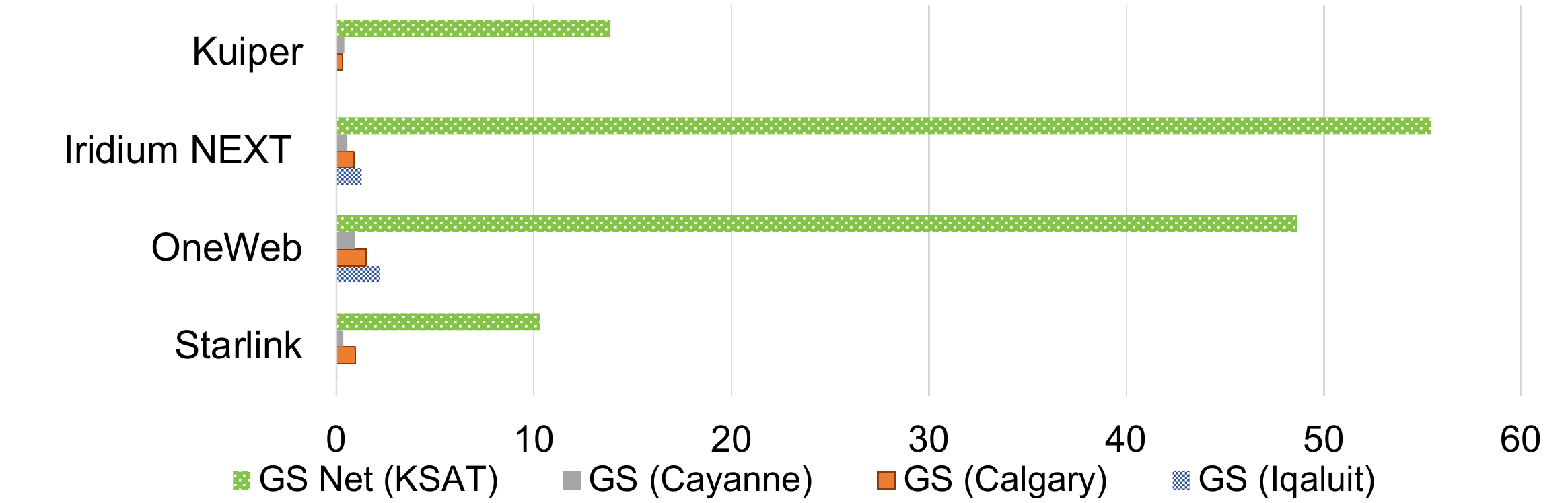}
    \vspace{-9pt}
    \caption{Per-satellite access ratio over a 24-hr mission in satellite constellations}
    \label{Fig:access_analysis}
    \vspace{-5pt}
\end{figure}

The traditional approach for satellite operations requires much manual process and relies on the accessibility between a ground station (GS) and target satellites for operational missions \cite{NASA2021} such as TT\&C performed through the ground facilities such as space/satellite operations centers.
With the launches of proliferated satellites in large and mega-constellations, the traditional operational process comes with significant challenges in efficiency. The efficiency issues can be seen from the access analysis results through our simulations with MATLAB shown in Table \ref{Tbl:access_analysis}, where three representative GS locations are considered in Northern Canada, Western Canada, and South America close to the equator. Our results list the access results for each constellation: ratio of accessible satellites by GS out of all satellites, the accessible time per accessible satellite $T_a$ (i.e., time for a satellite accessible by a GS), the ratio of access time per satellite over the entire mission duration $\gamma$. The orbits of the LEO satellites follow a Walker Star/Delta type of constellations \cite{Homssi2022}, where Telesat uses both types in its polar and inclined constellations, respectively, whose results are separated into two categories in Table \ref{Tbl:access_analysis}. Overall, Table \ref{Tbl:access_analysis} shows the Telesat constellation has the highest access time per accessible satellite, but its $\gamma$ value is very low, ranging from 1.489\% to 2.457\%. The results indicate the chance of accessing a single satellite from a GS facility for operations over a 24-hr mission is low, indicating a low possibility of performing an operational task meeting a stringent timing requirement in the \hl{traditional approach}. Even if a GS network such as the well-known KSAT GS network \cite{KSAT} in 26 sites is used, the per-satellite access ratio can only be improved to 10\% for Starlink and up to 55\% for Iridium NEXT as shown in Fig. \ref{Fig:access_analysis}.
Another important challenge found through the analysis is that $T_a$ or $\gamma$ largely bounds the amount of data transferred to/from a satellite. For example, if we pick the highest value of $T_a$ in Table \ref{Tbl:access_analysis}, $T_a = 2123.6$ s, the data amount transferred can achieve up to $T_a \cdot r$ where $r$ is the rate of a space-ground link. If we suppose a data rate of 1.8 Gbps on an optical link \cite{Zech15} is used, the data amount transferred through the Telesat constellation can only up be achieved to 3.8 Gb.

\subsection{Challenges in Satellite Communications and Networking}
One challenge in satellite communication and networking stems from dynamic demands and numerous resources with quality-of-service (QoS) guarantees.  
Another challenge arises from the network system complexity arising from the NGSO satellite constellations with high mobility and various hardware/software entities involved in missions, where the communication links and interfaces are subject to a variety of disruptions. 
These challenges leave satellites susceptible to a number of anomalies. Onboard and network anomalies on satellite subsystems and services can result in various types of space mission degradation or interruption. These anomalies can jeopardize the critical system's operation and pose significant economic losses. Anomalies can result from different causes. For instance, anomalies may come from the space environment, weather conditions, 
electrostatic discharge \cite{Choi2011}, operational errors, and malicious actions. The errors may cause malfunctioning inter-satellite link (ISL) issues for LEO satellite constellations. Weather conditions may cause interruptions or outages of the satellite-ground links. The failures on other network segments may result from device hardware failures on network nodes, such as processors, storage devices, power modules, and network interfaces. These failures occurring on a single compute node can contribute to outage events and affect connection and QoS metrics, such as the bit error rate, packet loss ratio, latency, data rate, and throughput. Various hardware/software-related issues can cause numerous network service anomalies. These challenges will significantly affect the quality of satellite communications and network services and have not been well addressed in the literature for satellite operations. Effective and intelligent solutions are required to handle the challenges in proliferated NGSO satellite networks.

\subsection{Challenges in Security, Resilience and Sustainability}
The prospects of fast-growing space assets come with critical security, resilience and sustainability challenges. For example, a communications process with limited ground-based infrastructure for data and TT\&C leaves the NGSO satellites prone to various anomalies resulting from onboard, network, and atmospheric issues. The intensive use of ISLs for data traffic between satellites may considerably expands the range of anomalous events and broadens the attack surface. 
Compared to the geostationary orbit (GEO) satellites, the shorter life span of small LEO satellites for about 5-10 years poses various sustainability issues for satellite disposal in the \textit{terminate} phase. 

The satellite conjunction risks can result in sustainability issues in operations. Although they are maintained at a low level analyzed in the \textit{design} phase, there are still conjunction risks from various space environments, objects, and anomalous events. A past collision between Iridium 33 and Kosmos 2251, for example, occurred due to insufficient avoidance maneuvers, considering the high frequency of close approaches of space vehicles, high velocity, and cost. Conjunction risks are considered significant threats in today's satellite constellations. A satellite orbital conjunction report on May 23, 2022, from the CelesTrak SOCRATES, shows the proximity of 8 m between Starlink 3890 and Starlink 3896. In addition, the space debris from satellite launches, on-orbit missions, and end-of-life decommissions will introduce risks to spacecraft safety and ruin the space environment.

Furthermore, future missions on advanced satellite systems are expected to support versatile communications and computing services. Enhancing the resilience and sustainability of NGSO satellites can help reduce service interruptions, degradation, capital expenditures (CAPEX), and operating expenses (OPEX). The revamped satellite systems call for an efficient operations approach to these challenges.

\begin{table*}[ht!]
\centering
\caption{Access analysis for commercial satellite constellations}
\label{Tbl:access_analysis}
\scalebox{0.76}{
\centering
\begin{tabular}{crrrrr}
\toprule 

\multicolumn{2}{r}{{\bfseries \shortstack{Constellation$^*$\\{} }}} &
{\bfseries \shortstack{Total \\Satellites}}&
{\bfseries \shortstack{Ratio of Accessible \\Satellites}} &
{\bfseries \shortstack{Access Time Per \\Accessible Sat. $T_a$ (s)}}&
{\bfseries \shortstack{Ratio of Access Time\\Per Satellite $\gamma$}}\\
\midrule
\multirow{5}{*}{GS1 (Iqaluit)} & {Starlink} & 1897 & 2.847\% & 24.576 & 0.028\% \\
& {Iridium NEXT} & 75 & 100\% & 1102.800 & 1.276\%  \\
& {Telesat (Polar)} & 78 & 100\% & 1875.800 & 2.170\% \\
& {Telesat (Inclined)} & 220 & 100\% & 588.682 & 0.681\%\\
& {Kuiper} & 3236 & 0 & 0 & 0  \\
\midrule
\multirow{5}{*}{GS2 (Calgary)} & {Starlink} & 1897 & 99.262\% & 831.871 & 0.962\% \\
& {Iridium NEXT} & 75 & 100\% & 760.800 & 0.880\% \\
& {Telesat (Polar)} & 220 & 100\% & 1218.800 & 1.410\% \\
& {Telesat (Inclined)} & 1897 & 100\% & 2123.600 & 2.457\% \\
& {Kuiper} & 3236 & 33.488\% & 247.233 & 0.317\% \\
\midrule
\multirow{5}{*}{GS3 (Cayenne)} & {Starlink} & 1897 & 98.050\% & 296.552 & 0.343\% \\
& {Iridium NEXT} & 75 & 98.667\% & 468.400 & 0.542\% \\
& {Telesat (Polar)} & 78 & 100\% & 725 & 0.839\% \\
& {Telesat (Inclined)} & 220 & 100\% & 1287.100 & 1.489\% \\
& {Kuiper} & 3236 & 100\% & 342.377 & 0.396\% \\
\bottomrule
\end{tabular}
}
\begin{itemize}\tiny{
\item[*]{The TLE data retrieved on May 11, 2022, are used for Starlink, Iridium NEXT, and OneWeb constellations. Telesat and Kuiper constellations are generated based on the the orbit propagator.}
\item[] {For consistency and fair comparison, the minimum elevation angle 25{\textdegree} is used for all constellations. A higher value of the evaluation angle may be used in the constellations such as 35{\textdegree} used for Kuiper. }
\item[]{The mission duration is 24 h. The results for Kuiper are averaged from its three orbital shells at the altitudes of 590 km, 610 km, 630 km.}
}
\end{itemize}
\vspace{-13pt}
\end{table*}

\begin{figure*}[ht]
    \centering
    \includegraphics[width=0.62\linewidth]{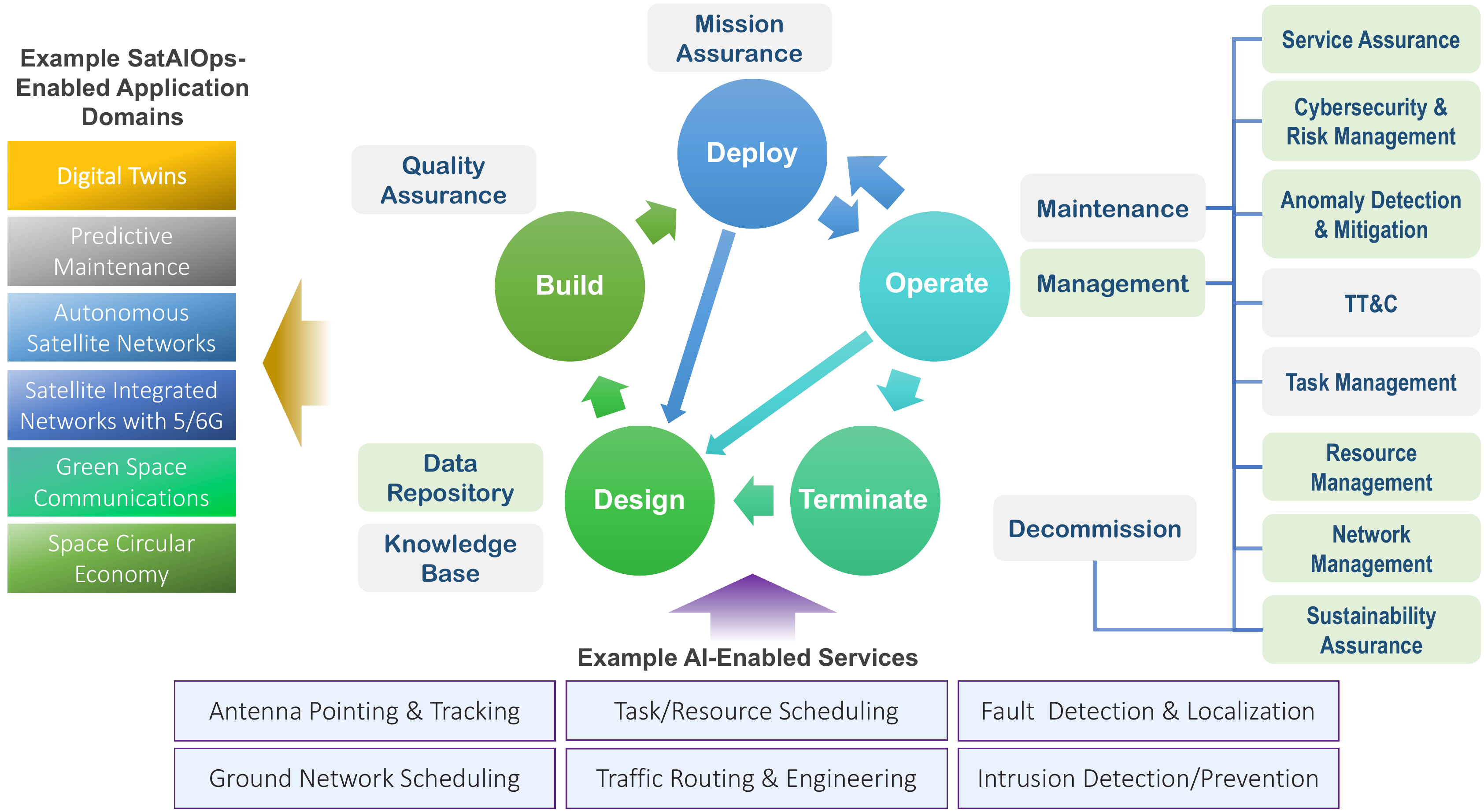}
    \vspace{-8pt}
    \caption{A SatAIOps process for a satellite's entire life cycle}
\label{Fig:SatAIOps_process}
\vspace{-12pt}
\end{figure*}

\begin{figure}[ht]
    \vspace{-8pt}
    \centering
    \includegraphics[width=0.66\linewidth]{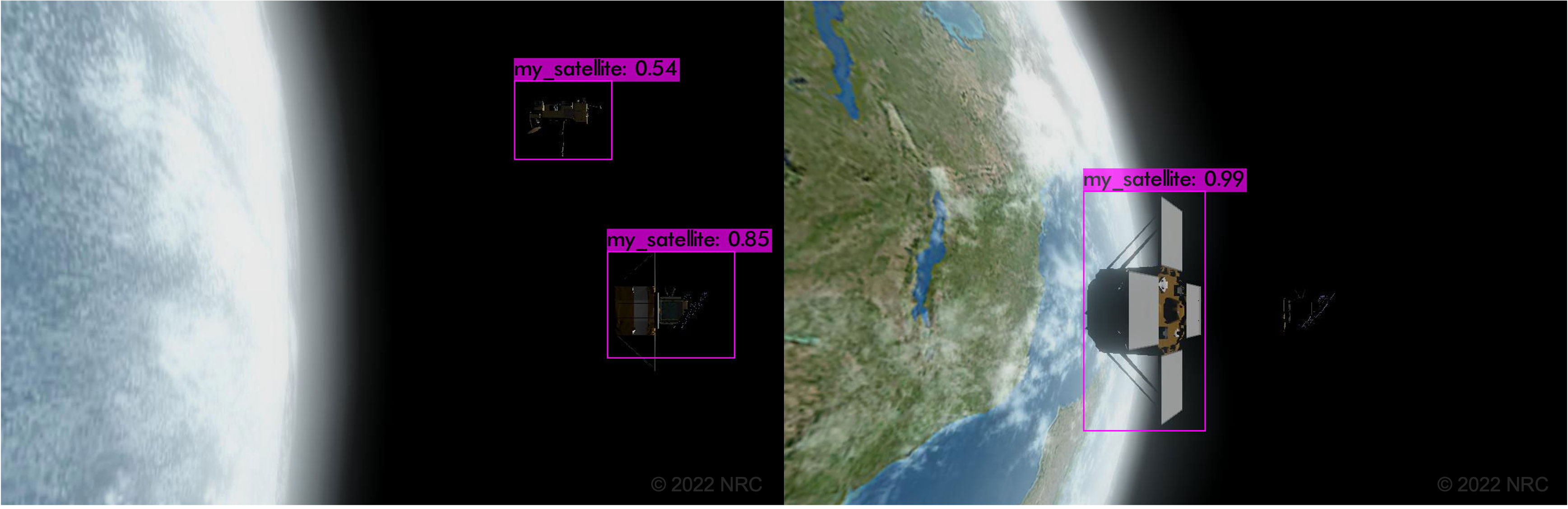}
    \vspace{-9pt}
    \caption{A demo of a proposed satellite conjunction monitoring service}
    \label{Fig:demo}
    \vspace{-18pt}
\end{figure}

\section{SatAIOps Approach}
To address the operational challenges in proliferated satellite networks, we propose a new approach, SatAIOps, using artificial intelligence (AI) for satellite network operations.

\subsection{Approach Overview}
Traditional satellite operations only address a few mission-related activities after deployment and view the life-cycle management in a uni-direction. The proposed SatAIOps aims to facilitate and improve the efficiency of the traditional operational approach. From the life-cycle perspective, SatAIOps enables the autonomous processes in the phases to enhance the efficiency and human/AI-in-the-loop interactions between different phases. For example, as shown in Fig. \ref{Fig:SatAIOps_process}, the \textit{design} phase can benefit from the \textit{deploy}, \textit{operate}, and \textit{terminate} phases. The \hl{data or models} obtained through non-design phases can help improve the design of a satellite system. They can also be utilized to assist the \hl{functional blocks} in the later phases, such as anomaly detection \& mitigation, mission \& service assurance, resource management, network management, and sustainability assurance. 

SatAIOps aims to achieve the following goals: (a) enhance the efficiency of satellite network operations, where the multi-layer networking (MLN) \cite{Hu_22_LATINCOM, Hu2023} can be used as a fundamental service; 
(b) utilize the softwarized and computing services to be running in space-based assets; (c) optimize the software and hardware development activities; and (d) adapt to the requirements for dynamic tasks and applications. 
In support of autonomous decisions on a space component, some fundamental AI services with reliable AI algorithmic execution and management need to be considered in SatAIOps. 

\subsection{SatAIOps Functions}
SatAIOps strives to tackle the challenges faced by the traditional operations approach with AI or machine learning (ML) based solutions. 
Compared to the traditional operational activities, additional functions are envisioned to be present in SatAIOps to address the complexity brought by the NGSO satellites in large and mega-constellations. With the advanced satellite platforms, \hl{AI/ML} methods can be employed in the space and ground components, including GSs and user terminals (UTs). The intelligent solutions in SatAIOps will replace some time-consuming or manual processes in operations. \hl{Trustworthy and responsible AI technologies} \cite{NIST_trustAI} will play a key role in these solutions. The main \hl{functional blocks} shown in Fig. \ref{Fig:SatAIOps_process} are discussed as follows.

\subsubsection{Mission Assurance} 
It includes the pointing and tracking of satellite antennas, the assurance of the satellite launch, and GS configuration for satellites. The tasks and missions can be dynamically performed with the \hl{reinforcement learning (RL)} methods. For example, the recurrent neural network (RNN) and deep reinforcement learning (DRL) techniques have been proposed to improve antenna pointing and tracking \cite{Xiao20} compared to a typical antenna search algorithm. 

\subsubsection{Data Repository}
It manages the essential datasets and AI models to be used in operations missions. Based on the relationship between the \textit{design} and \textit{deploy}, \textit{operate}, and \textit{terminate} phases, the data collected will be stored in raw or processed formats in the module to facilitate satellite system design and operations tasks. AI models are also managed in the module, which may be utilized in different operational tasks with transfer learning.

\subsubsection{Service Assurance} 
It provides quality service through satellite networks based on service-level agreements. The service level agreements (SLAs) need to be dynamically met through trustworthy AI-based algorithms, such as DL and deep learning (DL).

\subsubsection{Cybersecurity \& Risk Management and Anomaly Detection \& Mitigation} The SatAIOps solutions need to handle the threats and risks from, for example, cyberattacks, onboard/network anomalies, and collisions. Autonomous/autonomic networking \cite{Hu22, rfc7575} can help with the countermeasures to the threats and risks. To mitigate the threats of cyberattacks, reliable, efficient, and intelligent countermeasures to attacks are expected. 
Multivariate time-series analysis can be used to handle anomaly detection with telemetry or diagnostic data at the onboard or network scale. For example, a long short-term memory (LSTM) prediction method for unlabeled spacecraft data has been implemented in \cite{hundman18}. For labelled data, the ML classification algorithms can be used to provide insight into the root causes for mitigation, where mitigation schemes can be realized through space, aerial, and terrestrial network components \cite{Hu22}. In another example, satellite networks can be deployed with AI algorithms to process the telemetry and traffic data for threat/fault detection \cite{Hu_22_TAES}, fault localization, and mitigation measures. Threats and risks to be handled with this function may come from cyberattacks, onboard and network anomalies, collisions, and security risks in the supply chain in a \textit{design} or \textit{build} phase \cite{Falco2018}. 

\subsubsection{Resource Management} 
It manages the platform, payloads, communication (e.g., radio-frequency (RF) and optical wavelength bands/beams, software-defined radio (SDR) modules, etc.), and compute resources (e.g., onboard and cloud/edge computation entities) in a spacecraft fleet. This includes optimal resource scheduling and utilization in a spacecraft and over the network in a pre-defined or dynamic fashion. AI/ML models deployed on the satellites can also be utilized by the function through an interface with the data repository function. An example application is the SDR \cite{TORRIERI2021} with AI algorithms on the satellites that can be adaptive to ISL and space-ground link conditions. 

\subsubsection{Network Management} 
It manages the satellite networks and services arising from the advanced satellite platforms. The regenerative satellites are expected to have onboard processing capability, and software-defined networking (SDN) has been proposed to have space components in the satellite-terrestrial integrated networks. The components in this network management function may include ground and space entities and include general functionalities, such as network administration, flow monitoring/control, routing, virtual network functions (VNF), and QoS-guaranteed network services. The SDN controllers can be optimized in an example scenario through the adaptive controller placement problem formulated \cite{9488806}. The RL schemes can benefit an optimal path through ISLs and the orchestration of satellite compute resources. 

\subsubsection{Sustainability Assurance}
It can help achieve the goals of environmental and service sustainability from different satellite missions. The management function will ensure the contingency and environmental effects in the \textit{operate} phase of the satellite's life cycle. Decommission sequences need to guarantee minimal environmental impact in the \textit{terminate} phase, where data generated in this phase can be used to optimize the activities in the \textit{design} phase. For example, an onboard AI method may perform the conjunction-risk avoidance task through an automated manoeuvre. Satellites can also be equipped with vision-based on-orbit servicing to monitor conjunction risks in real time, such as our work demonstrated in Fig. \ref{Fig:demo}, and execute the propulsion system for collision avoidance. This AI-enabled solution may be applied to the decommission sequence to mitigate space debris generation.

\subsection{AI-Enabled Services} 
As shown in Fig. \ref{Fig:SatAIOps_process}, underlying AI-enabled services can be used for SatAIOps functions. For example, the antenna pointing \& tracking and ground network scheduling services can leverage the AI/ML algorithms for a wide range of tasks for SatAIOps functions. The task/resource scheduling service can be dynamically performed with RL methods. The fault detection \& localization service can provide a foundation for AI/ML solutions consuming various sources of data, such as telemetry, traffic, log files, and trace data. The intrusion detection/prevention service can handle cyber threats and attacks. The traffic routing \& engineering service can provide reliable communication and optimal scheduling of dynamic tasks and applications. For example, an MLN-based scheme \cite{Hu_22_LATINCOM, Hu2023} can assist with TT\&C and other operational missions. Through transfer learning, AI solutions for handling different missions may be generalized to different use cases. 

\subsection{SatAIOps-Enabled Application Domains}
The proposed SatAIOps can enable several application domains based in Fig. \ref{Fig:SatAIOps_process}. For instance, a digital twin of a satellite node can enable its operational status and services in real time. SatAIOps can provide essential services integrated into digital twins to facilitate the activities in all life-cycle phases. Predictive maintenance can check and predict the condition of space and ground components, aligned with the SatAIOps goals in the \textit{operate} phase. GSs can employ AI algorithms for optimal RF and free-space optical communications with space components. A UT may require AI models for a beamforming approach in a multiple-input and multiple-output (MIMO) scenario, such as single-user, multi-user, and massive MIMO \cite{You20_MIMO}. A predictive maintenance operation may be done through in-orbit maintenance missions discussed in \cite{KRISHN2016}. Currently, some predictive maintenance tools have reportedly been developed for ground and spacecraft health \cite{Harebottle2019}. Autonomous satellite networks represent a promising domain for autonomous network and service management for satellite networks. SatAIOps can foster green space communications, space circular economy \cite{ESA_cleanspace}, and integration with terrestrial telecommunications networks, where non-terrestrial network (NTN) is being standardized in the evolution of beyond 5G networks. 

\section{Conclusion}
This work discusses new perspectives to address the emerging challenges resulting from the fast-growing NGSO satellites in large and mega-constellations. The proposed SatAIOps approach handles reliable and efficient solutions to critical tasks and missions throughout the satellite's entire life cycle, leading to enhanced resilience, sustainability, mission assurance, security, and network management that can be handled in a holistic manner. SatAIOps also promotes trustworthy and responsible actions in all operational aspects and various initiatives such as autonomic networking/management and ESA's Clean Space initiatives. 

 \section*{Acknowledgment}
We acknowledge the support of the High-Throughput and Secure Networks Challenge program of National Research Council Canada, and the support of the Natural Sciences and Engineering Research Council of Canada (NSERC), [funding reference number RGPIN-2022-03364].


\bibliographystyle{IEEEtran}

\bibliography{./bibitems}






\vfill

\end{document}